\documentclass[aps,twocolumn]{revtex4}
\usepackage{graphicx}
\begin{document}

\title{Nanometer-spaced platinum electrodes with calibrated separation}
\author{Y.-V. Kervennic, H.S.J. Van der Zant, A. F. Morpurgo, L. Gurevich, and L. P. Kouwenhoven}
\affiliation{Kavli Institute of Nanoscience, Faculty of Applied
Sciences, Delft University of Technology, Lorentzweg 1, 2628 CJ
Delft, the Netherlands}

\begin{abstract}

We have fabricated pairs of platinum electrodes with separation between 20 and 3.5~nm. Our technique combines electron beam lithography and chemical
electrodeposition. We show that the measurement of the conductance between
the two electrodes through the electrolyte provides an accurate and
reproducible way to control their separation. We have tested the robustness
of the electrodes by applying large voltages across them and by using them
to measure the transport properties of Au nano-clusters. Our results show
that the technique reliably produces metallic electrodes with a separation
that bridges the minimum scale accessible by electron beam lithography with
the atomic scale.
\end{abstract}

\maketitle

\begin{figure}[htbp]
\centering
\includegraphics[width=9cm]{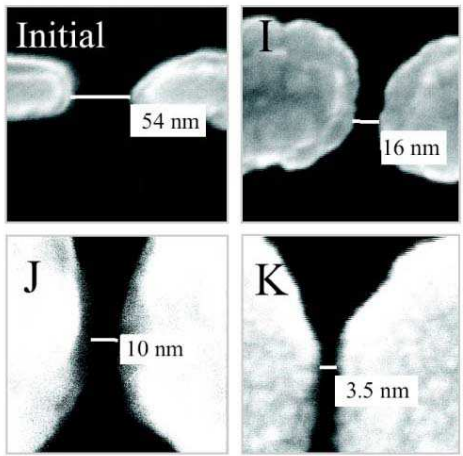}
    \caption{Upper left corner: the initial separation of the electrodes as defined by electron beam lithography and lift-off is 60~nm. Figures I, J and K show electrodes fabricated by stopping the electrodeposition process when the monitor current was 30, 90 and 140~nA respectively. The electrode separations corresponding to these values are 16, 10 and 3.5~nm.}
    \label{figch41}
  \vspace{-0.5cm}
\end{figure}

Modern chemistry has produced a large variety of objects perfectly defined
on a molecular level that are of interest for the development of nano-scale
electronics~\cite{joachim00}. Examples are metallic and semiconducting clusters,
conjugated organic molecules and fullerenes. The investigation of their
individual properties and their use in practical circuits requires the
ability to interface these objects with the macroscopic world. In particular,
to perform electrical measurements, contacts must be made to single
nano-objects. Over the past few years ingenious methods have been
introduced to fabricate metallic leads with a separation on the nanometer
scale~\cite{park,klein,kergueris,reed}. 
However, a simple, fully controllable and reproducible 
technique capable of bridging the dimension scale accessible by
(e)-beam lithography ($\sim$~20-50~nm) with the atomic scale 
($\sim$~0.1-1~nm) is still lacking.\\
\indent Recently, a technique (hereafter referred to as \textquoteleft electrochemical narrowing\textquoteright) combining conventional lithography and electrochemistry has
allowed the fabrication of metallic electrodes with atomic separation~\cite{morpurgo,li}. 
In this technique, metal electrodeposition on two planar
electrodes (fabricated by conventional means) is used to reduce their
separation. {\em In-situ} monitoring of the conductance
between the two electrodes provides
the desired control. The simplicity and robustness of the process as well as
its flexibility (e.g., electrodes fabricated in this way can be readily
integrated with electrostatic gates) are some of its key aspects.\\
 \indent So far, only the working principle of electrochemical narrowing has been demonstrated but many important issues remain to be solved. In particular, no
optimization of the process and not even of the metal to be deposited have been performed yet. Also, and
more fundamentally, it is unclear whether the monitoring of the conductance
can be used to control the inter-electrode separation on a length scale
larger than 1-2 nanometers and, if so, whether it is possible to calibrate
this control method with sufficient accuracy and reproducibility.\\
\indent Here, we report a systematic study of nano-electrode fabrication by means of
electrochemical narrowing. We show that by optimizing the electrode layout
and the control circuit we can monitor the inter-electrode conductance for
separations up to 20 nm. By stopping the electrodeposition process at
predefined conductance values we reproducibly obtain the same
electrode separation with a $\sim$~10\% error. We have found that
the use of Pt allows for high reproducibility and high stability of the
electrodes (in comparison to deposition with Au). 
We have tested the quality of the electrodes by performing
electrical measurements on individual Au nano-clusters with a diameter
ranging from 20 to 5~nm.\\
\indent Details of the electrode fabrication are as follows. With e-beam lithography, 
we define the \textquoteleft large\textquoteright~separation electrodes on thermally oxidized silicon wafers by evaporating 7~nm of Ti and 25~nm of Pt. After lift-off, the
initial separation is between 40 and 80~nm (see Fig$.$~\ref{figch41}). The samples are then
dipped in a buffered solution of hydrofluoric acid (AF 87.5-12.5, LSI Selectipur, Merck) for 20
seconds which etches approximately 30~nm of SiO$_2$. As a result,
the extremities of the electrodes are free-standing. Such
free-standing contacts reduce the possible channels of parallel conduction on
the substrate surface. The sample layout away from the electrode
extremities is also of importance. As compared to Ref.~\cite{morpurgo} 
we have reduced the electrode area that is in contact with the electrolyte, which improved the monitoring sensitivity.\\
\indent Electrodeposition of Pt is performed in an aqueous solution of 0.1~mole of
K$_2$PtCl$_{4}$ and 0.5~moles 
of H$_2$SO$_4$~\cite{lin}. Contrary to gold, the quality of
electrodeposited Pt is very sensitive to the initial conditions of the
electrode surface. Gold can readily be plated from a cyanide bath containing
gold ions without any pre-treatment of the electrodes. When following the
same procedure with Pt, the material grows with a very rough morphology
(i.e., grains of random size nucleates at random locations). This morphology
does not allow the fabrication of nanoscale contacts.\\
\indent To obtain a
smooth Pt surface we first clean the evaporated platinum electrodes in an
O$_2$ plasma. We then put the electrodes
in solution and apply a square wave signal with respect to the counter
electrode for a few seconds. The voltage changes from - 1.8~V to + 0.7~V in periods of 400~ms for a few seconds. Pt is then electrodeposited by biasing the two electrodes at 420~mV
with respect to the counter-electrode (a Pt wire).
This process produces uniform Pt grains, whose size is less
than 6~nm. Such a small grain size is normally not obtained with Au.

\begin{figure}[htbp]
\centering
\includegraphics[angle=0,width=9cm]{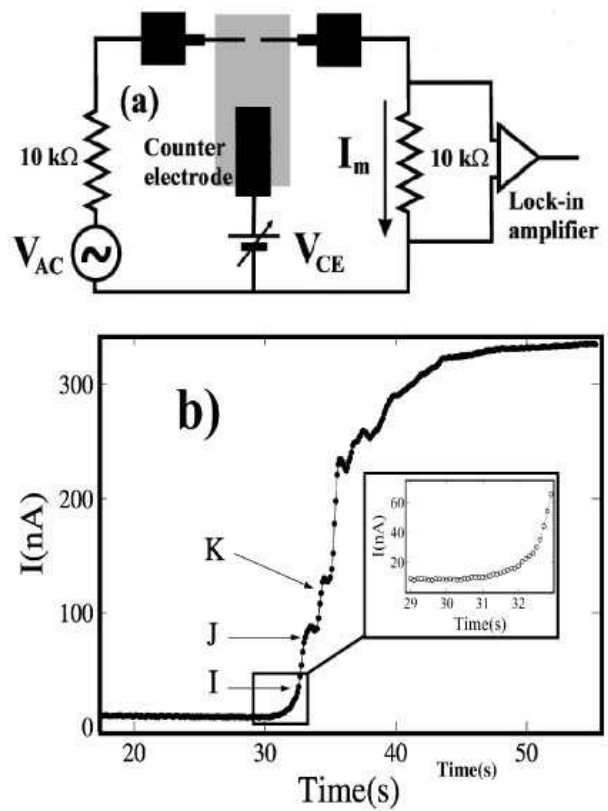}
\vspace{-1.5cm}
    \caption{a) Circuit schematics to electrodeposit Pt and to
monitor the electrode separation. During the electrodeposition process, the
voltage applied to the counter-electrode is $V_{\rm CE} = 420$~mV. The monitor
current is measured in phase with the ac excitation using a lock-in amplifier ($V_{\rm ac} =10$~mV, $f=10$~Hz). The shaded region represents the part of the circuit immersed in the electrolyte. b) Monitor current versus time. A pronounced increase of $I_m$ is observed as the electrode separation decreases (the inset shows the initial phase of this increase). At three predetermined values of the
monitor current (pointed by the arrows) electrodeposition was stopped and
the electrode distance measured using a SEM (see
Fig$.$~\ref{figch41}).}
    \label{figch42}
  \vspace{0cm}
\end{figure}

Monitoring the inter-electrode conductance during growth is done using a
lock-in amplifier (see Fig$.$~\ref{figch42}a). Compared to Ref.~\cite{morpurgo} the monitoring circuit has been improved. First, the resistor used to measure the
ac monitor current $I_m$ is
larger (10 k$\Omega$) which gives higher sensitivity at small currents. Also, an additional 10 k$\Omega$ resistor has been added to make the electroplating process more
symmetrical. A typical trace of $I_m$ vs. plating time is shown in Fig$.$~\ref{figch42}b. In this particular case the gap has been closed to show the
full behavior of $I_m$. Low $I_m$ on the left side corresponds to a large electrode separation. At the right side (high current), the electrodes are completely bridged with electrodeposited platinum. 
The curve contains step-like features. Scanning Electron Microscope (SEM) imaging on many sample samples suggest that they are caused by the granular Pt growth.

To calibrate the relation between monitor current and electrode
separation we stop the electrodeposition process at different, predetermined
values of $I_m$. The electrodes are then removed from the solution, rinsed with
de-ionized water and blown dry with nitrogen. After cleaning in an O$_2$
plasma, samples are mounted in a Hitachi s-900 Scanning Electron Microscope which has a nominal resolution better than 1~nm. This allows a precise determination of the electrode
separation. We find that to a given value of $I_m$ corresponds to a
well-determined value of the electrode separation, which is
reproducible within $\sim$~10\%~\cite{comment2}. For
the geometry and electrolyte discussed in this chapter we have fabricated
and carefully inspected more than 20 electrode pairs: we find that
$I_{m} = 30$~nA corresponds to a separation of 16~nm, $I_m= 90$~nA to 10~nm and $I_m= 140$~nA to 3.5~nm.
\begin{figure}[htbp]
\centering
\includegraphics[angle=0,width=5cm]{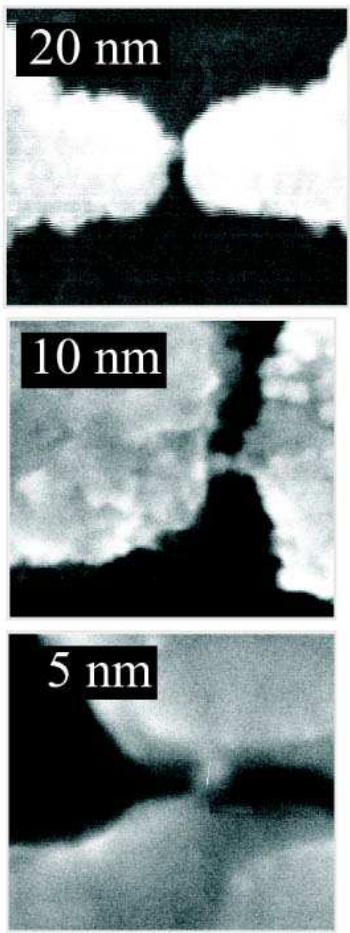}
      \caption{SEM micrographs of three electrode pairs of different separation with individual Au clusters trapped in between them. From top to bottom, the cluster size is 20~nm, 10~nm and 5~nm. No cleaning prior to SEM imaging is possible without damaging the sample. This explains why the images of these samples are less sharp than those shown in Fig$.$~\ref{figch41}.}
  \label{Elec}
\end{figure} 
 
This reproducible relation between $I_m$ and the electrode
separation, extending to separations as large as 20~nm, has not been
reported in previous experiments and is one of the main results of this
chapter. It demonstrates that electrochemical narrowing can be controlled and
used to bridge the dimension scale accessible with e-beam lithography to the
atomic scale. We believe that the ability to electrodeposit Pt with very small grains is required to achieve the reproducibility and
controllability reported here~\cite{comment3}.

The behavior of the monitor current as a function of electrode separation
is worth a comment. It has been speculated~\cite{morpurgo} that the large current
flowing between the electrodes at small separation is due to direct electron
tunneling, enhanced by the presence of ions in the electrolyte. 
However, it seems unlikely that a considerable contribution to the current can
originate from direct tunneling for separations exceeding 20~nm. We presently
do not understand the precise mechanism behind the increase of $I_m$ at small separation.\\
\indent Pt electrodes with a few nanometer separation are remarkably robust. We have applied up to 5~V in air across electrodes separated by 3.5~nm, without
observing any sign of degradation. This is in marked contrast with  Au electrodes. For a comparable separation,
Au electrodes normally break down when a few hundred millivolts bias
is applied. This robustness makes Pt electrodes suitable to measure small
nano-objects at high electric field. 
\begin{figure}[htbp]
\centering
\includegraphics[width=9cm]{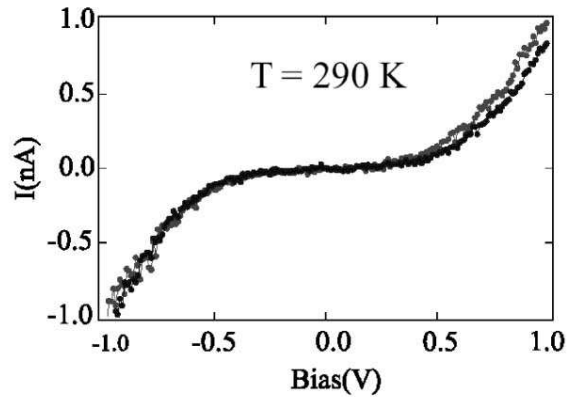}
  \vspace{-0.5cm}
    \caption{Two room-temperature $I$-$V$ characteristics of a 5~nm cluster trapped between Pt electrodes, taken a few minutes after each other. The $I$-$V$ curves display a flat region of about 500~mV. This value corresponds well with the one expected from the theory of coulomb blockade ($2E_C=580$~meV).}
    \label{figch4}
  \vspace{-0.5cm}
\end{figure}
As an example, we have measured the
electrical characteristics of individual Au nano-clusters of different
dimensions (20~nm, 10~nm and 5~nm, commercially available from BBI
international in mono-disperse solutions), electrostatically trapped~\cite{bezryadin} (see Fig$.$~\ref{Elec}).
The room-temperature resistance of trapped clusters typically ranges between
100~ M$\Omega$ and 10~G$\Omega$, whereas the resistance across
the electrodes before trapping is always larger than 100~G$\Omega$, which is the limit of our measuring apparatus. These variations in
the resistance can easily be explained by different geometrical
contact configurations between the electrodes and the clusters, or by a
different level of oxidation of the Pt surface.\\ 
\indent Figure~\ref{figch4} shows $I-V$ curves measured at room temperature on a 5~nm cluster. It exhibits a pronounced non-linearity. The current suppression observed at low bias is consistent with Coulomb blockade through the cluster. The estimated self-capacitance of the cluster $C_0 =0.27$~aF gives a charging energy of $E_C=290$~mV, which compares well to the size of the measured gap. 
To demonstrate the reproducibility of the measurements, two curves are shown in Fig$.$~\ref{figch4}.\\ 
\\
\indent
We acknowledge Prof. Klapwijk's group for kindly hosting our set up, Danny
Porath for sharing information, Anja Langen, Bert de Groot and Raymond
Schouten for technical advice, Daniel Van Maekelbergh for discussions, Monica Monteza Farfan, Paul
Casas, Gilles Pirio and Adrian Bachtold for their support. This work is financed
by FOM and ERATO. H.~S.~J.~vdZ is supported by the Dutch Royal
Academy of Arts and Sciences (KNAW).

\vspace{0.5cm}

\noindent 

\begin{thebibliography}{99}
\bibitem{comment1a}
It is believed that this surface preparation procedure results in the
formation of a uniform distribution of nucleation centers for the growth of
electrodeposited Pt.
\bibitem{comment2}
For the smallest separation the error is somewhat larger, limited by
the SEM resolution $\sim$~1nm.
\bibitem{comment3}
The small grain size is essential to reliably determine the electrode separation. For electrochemically overgrown Au
electrodes, large grains often make an accurate determination  impossible.
\end{thebibliography}

\end{document}